\documentclass[%
 aip,
 amsmath,amssymb,
 reprint,%
]{revtex4-1}

\usepackage{graphicx}
\usepackage{dcolumn}
\usepackage{bm}
\usepackage{color}
\usepackage[utf8]{inputenc}
\usepackage[T1]{fontenc}
\usepackage{mathptmx}
\usepackage[driverfallback=dvipdfmx,colorlinks,linkcolor=blue,citecolor=blue,urlcolor=blue]{hyperref}

\begin{document}

\title{Imaging the dipole scattering of an optically levitated dielectric nanoparticle}

\author{Yuanbin Jin}
 \affiliation{State Key Laboratory of Quantum Optics and Quantum Optics Devices,Institute of Opto-Electronics, Synergetic Innovation Center of Extreme Optics, Shanxi University, Taiyuan 030006, P.R.China}
\author{Jiangwei Yan}
 \affiliation{State Key Laboratory of Quantum Optics and Quantum Optics Devices,Institute of Opto-Electronics, Synergetic Innovation Center of Extreme Optics, Shanxi University, Taiyuan 030006, P.R.China}
\author{Shah Jee Rahman}
 \affiliation{State Key Laboratory of Quantum Optics and Quantum Optics Devices,Institute of Opto-Electronics, Synergetic Innovation Center of Extreme Optics, Shanxi University, Taiyuan 030006, P.R.China}
\author{Xudong Yu}
 \affiliation{State Key Laboratory of Quantum Optics and Quantum Optics Devices,Institute of Opto-Electronics, Synergetic Innovation Center of Extreme Optics, Shanxi University, Taiyuan 030006, P.R.China}
\author{Jing Zhang}
 \email{jzhang74@sxu.edu.cn, jzhang74@yahoo.com}
\affiliation{State Key Laboratory of Quantum Optics and Quantum Optics Devices,Institute of Opto-Electronics, Synergetic Innovation Center of Extreme Optics, Shanxi University, Taiyuan 030006, P.R.China}

\begin{abstract}

We experimentally observe the dipole scattering of a nanoparticle using a high numerical aperture (NA) imaging system. The optically levitated nanoparticle provides an environment free of particle-substrate interaction. We illuminate the silica nanoparticle in vacuum with a 532 nm laser beam orthogonally to the propagation direction of the 1064 nm trapping laser beam strongly focused by the same high NA objective used to collect the scattering, which results in a dark background and high signal-noise ratio. The dipole orientations of the nanoparticle induced by the linear polarization of the incident laser are studied by measuring the scattering light distribution in the image and the Fourier space (k-space) as we rotate the illuminating light polarization. The polarization vortex (vector beam) is observed for the special case, when the dipole orientation of the nanoparticle is aligned along the optical axis of the microscope objective. Our work offers an important platform for studying the scattering anisotropy with Kerker conditions.

\end{abstract}

\maketitle

Optical tweezers using light field radiation pressure have greatly progressed after Ashkin's pioneering experiment \cite{PhysRevLett.24.156, doi:10.1063/1.88748}. Optically levitated nanoparticle for its high Q-factor \cite{doi:10.1142/S0217979213300181, Millen_2020} in vacuum can be used for weak forces detection and precise measurement \cite{PhysRevLett.105.101101, PhysRevA.98.053831, PhysRevLett.121.063602, doi:10.1063/1.5081045}. To reach the quantum ground state, many techniques for cooling the center-of-mass (CoM) motion of the trapped nanoparticle have been developed, such as speed feedback cooling \cite{NatPhys.7.527}, parametric feedback cooling \cite{PhysRevLett.109.103603, Vovrosh:s, doi:10.1063/1.5099284}, electric feedback cooling \cite{PhysRevA.99.051401, PhysRevLett.122.223601, PhysRevLett.122.223602}, and cavity-assisting cooling \cite{PhysRevLett.122.123601, PhysRevLett.122.123602, Deli892}. Recently, the temperature of the CoM motion of a nanoparticle has been cooled to the quantum ground state \cite{Deli892}, which can be used to research macroscopic quantum phenomenon such as the Schrodinger's cat state \cite{Romero_Isart_2010, Romero-Isart11, Wan16}. Moreover, the optically levitated nanoparticle systems have been studied for measuring the Casimir force and torque \cite{PhysRevLett.121.033603,PhysRevA.96.033843}, and GHz scale rotation was recently observed \cite{PhysRevLett.121.033603, PhysRevLett.121.033602, Li2020, jin20206}, including the record-breaking ultra-high rotation frequency of about 6 GHz by our group \cite{jin20206}.

Light scattering from the nanoparticles is an important physical process and plays an important role in a wide variety of applications such as surface enhanced spectroscopy, sensing and in the study of non-linear phenomena \cite{RevModPhys.57.783, PhysRevApplied.6.014018, PhysRevApplied.10.054005, PhysRevApplied.12.014010}. Recently, it has attracted an increased attention in the study of nanoantennas which modify their radiative characteristics and radiation pattern \cite{ncomms2538, PhysRevLett.102.146807, PhysRevLett.104.026802, ncomms8915, srep14788}. For example, based on Mie theory, the interference of electric and magnetic dipole resonances inside the high-refractive index dielectric nanoparticles can strongly affect their scattering pattern, which depends on the light wavelength, giving rise to interesting phenomena such as Kerker's-type scattering with zero backscattering \cite{ncomms2538}. Light scattering studies have usually been performed in metasurfaces or nanoparticles on a substrate, in which light scattering is modified in the presence of the substrate. It is worth noting that the dipole orientations of the nanoparticle induced by the linear polarization of the incident laser is different from the orientation of the geometrical shape of the nanoparticle. The orientation of an optically levitated nonspherical nanoparticle can be determined according to its dynamical properties \cite{PhysRevLett.121.033603,Stefan17}.

3D dipole orientations of a single-molecule have been extensively studied by the polarization analysis of the emitted fluorescence \cite{Fourkas:01, nc104694, PhysRevX.4.021037}, or by the recording of the defocused \cite{SEPIOL1997444, doi:10.1021/jp993364q, PhysRevLett.115.173002} or aberrated \cite{doi:10.1021/jp9846330} or k-space \cite{Lieb:04, Dodson:14, srep59966, nc115307} fluorescence image of a molecule. In these schemes, the excitation light is illuminated confocally along the optical axis of the microscope objective. In this paper, we report the imaging of the dipole scattering orientations of an optically levitated dielectric nanoparticle in the image and Fourier space (k-space). The main differences of our scheme from the other previous works are as follows: (i) Our setup provides an environment free of the particle-substrate interaction. (ii) We illuminate the nanoparticle with a excitation laser beam orthogonally to the optical axis of the microscope objective. This configuration provides a dark background and high signal-noise ratio for detecting the dipole scattering. (iii) Due to the configuration discussed above, the polarization vortex (vector beam) is observed when the dipole orientation of the nanoparticle is aligned exactly along the optical axis of the microscope objective.

\begin{figure}
\includegraphics[width=3in]{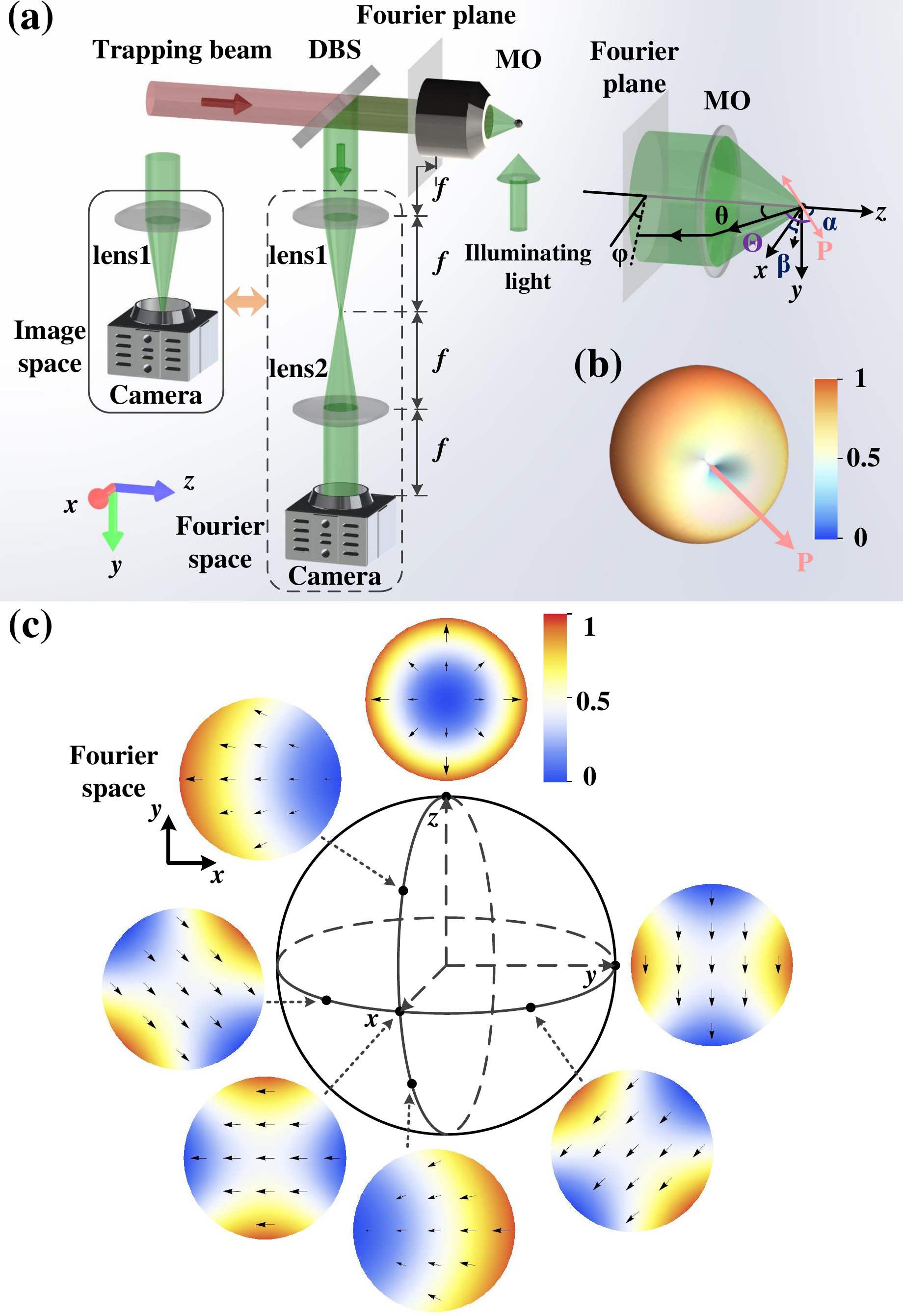}
\caption{Schematic diagram of the dipole scattering and imaging system. (a) The layout of the trapping and image systems. The dipole scattering can be measured in the image plane (the schematic diagram in solid box) and Fourier space (the schematic diagram in dashed box). (b) The spatial distribution of the dipole scattering in free space. (c) The dipole orientations (black spots) plotted on a three-dimensional Bloch sphere and the corresponding intensity distributions and polarization in k-space. MO: microscope objective; DBS: dichroic beam splitter.}
\label{fig1}
\end{figure}

A nanoparticle optically trapped by a strongly focused laser beam exhibits a CoM motion in the focus region. It can be regarded as a three-dimensional harmonic oscillator with a displacement of less than about one hundred nanometers. Another laser beam (with wavelength $\lambda$ larger than the scale of the nanoparticle) illuminates the nanoparticle. The scattering from the nanoparticle contains the all multipole terms, but only the dominant dipole term is considered here due to the low refractive index of the nanoparticle. Therefore, the spatial distribution of light scattering in free space, as a dipole radiation is given as
\begin{eqnarray}
{I} = \frac{{3}}{{8\pi r^2}}{I_{total}}{\sin ^2}\Theta
\label{eq:one},
\end{eqnarray}
where ${I_{total}} = {{\omega^4 \left| {\bf {p}} \right|}^2}/({12\pi {\varepsilon _0}{c^3}})$ is total radiation power at the angular frequency $\omega$, the electric dipole moment \textbf{p}, and the scattering field vector \textbf{r}. The parameter $\Theta$ is the angle between \textbf{p} (dipole axis) and \textbf{r}. In Cartesian coordinates, the angle of dipole axis (scattering field vector) relative to $z$-axis is $\alpha$ ($\theta$), and the azimuthal angle in $xy$-plane relative to $x$-axis is $\beta$ ($\varphi$). The distribution of the dipole scattering in free space can be written as

\begin{align}
&I(\theta,\varphi;\alpha,\beta)= \notag \\
&\frac{3}{{8\pi r^2}}{I_{total}}\left[ {1 - {{\left( {\cos \theta \cos \alpha  + \sin \theta \sin \alpha \cos \left( {\varphi  - \beta } \right)} \right)}^2}} \right]
\label{eq:two},
\end{align}

The dipole scattered light is collected by the same microscope objective, and the intensity distribution in the back-aperture plane of the microscope objective can be written as \cite{Lieb:04}
\begin{eqnarray}
I^{F}(\theta, \varphi; \alpha, \beta) \propto \frac{1}{{\cos \theta }}\left( {{{\left| {{e_p}} \right|}^2} + {{\left| {{e_s}} \right|}^2}} \right)
\label{eq:three},
\end{eqnarray}
where $e_p$ and $e_s$ are the p- and s-polarized components of electric field and are given by ${e_p} = \cos \alpha \sin \theta  - \sin \alpha \cos \theta \cos \left( {\varphi  - \beta } \right)$ and ${e_s} = - \sin \alpha \sin \left( {\varphi  - \beta } \right)$. Here, the standard apodization factor $1/\cos \theta$ is introduced in order to conserve the energy along each geometric path. The NA of the microscope objective determines the range of the angle $\theta$. The intensity distribution in the back-aperture plane of the microscope objective is the Fourier space (k-space) distribution of the scattering light. Figure 1(b) shows that the spatial distribution of the scattered light in free space resembles a 3D doughnut shape and the dipole axis passes through the hole of the doughnut-shape. Figure 1(c) shows the dipole orientations ($\alpha,\beta$) plotted on a three-dimensional Bloch sphere and the corresponding intensity distributions in k-space, which clearly illustrates the corresponding relationship between the dipole orientation and the image in k-space. In this experiment, the illuminating light propagates along the y-axis, and the induce dipole axis is aligned with the linear polarization of the illuminating laser. Thus the dipole axis is rotated along the meridian of the Bloch sphere by rotating this polarization (we change $\alpha$, but $\beta=0$).

\begin{figure}
\includegraphics[width=3.3in]{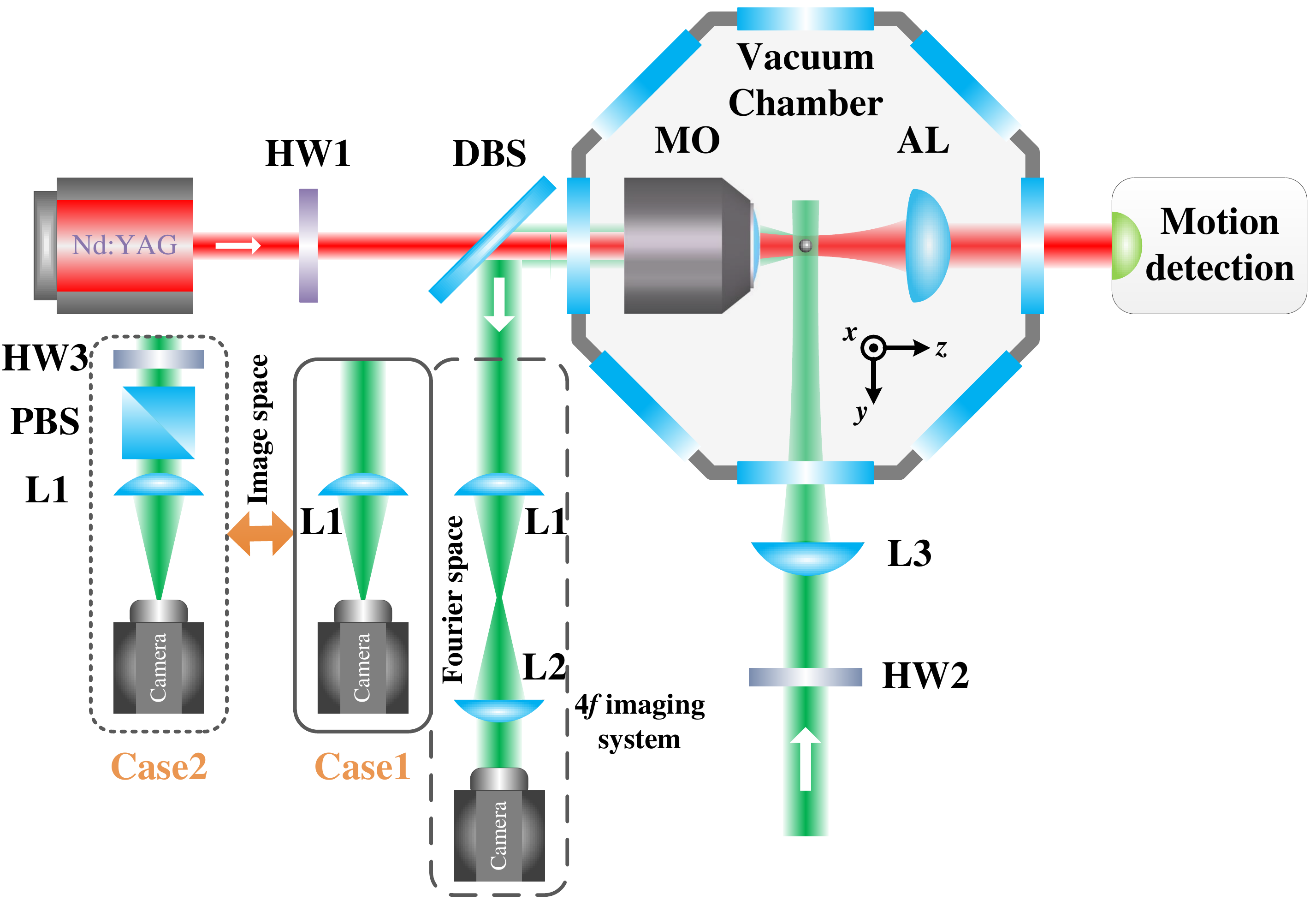}
\caption{Schematic diagram of the experimental setup. The 1064 nm laser beam is tightly focused by an microscope objective (NA=0.95). The 532 nm laser beam illuminates the nanoparticle, and the scattered light is collected by the same microscope objective into a camera. The right part is the detection scheme for measuring the motion signal of the nanoparticle. MO: microscope objective; AL: aspheric lens; DBS: dichroic beam splitter; PBS: polarization beam splitter; HW1-3: half-wave plates; L1-3: spherical lens.}
\label{fig2}
\end{figure}

In our experiment, we optically trap a silica nanoparticle in a vacuum chamber. The schematic diagram of the experimental setup is shown in Fig. 2, which is described in detail in Refs. 48 and 49. A 1064 nm laser beam generated by a single-frequency laser is tightly focused by a high NA microscope objective (NA = 0.95). The propagation direction of trapping beam is along $z$-axis. We collect the output light with another high NA aspheric lens (NA = 0.68) to detect the CoM motion signal of the nanoparticle using a balanced detection system. In order to load the nanoparticles, the commercial non-functionalized silica nanopaticles (TianJin Baseline) are first dissolved in ethanol, followed by 15 min of sonication. The nanoparticles are then diluted and poured into an ultrasonic nebulizer. The droplets containing the nanoparticles are dispersed by the ultrasonic nebulizer and guided by a long pipe into the focus of the microscope objective in the vacuum chamber. Once a particle is trapped, a vacuum pump is used to evacuate the chamber. The radius of the nanoparticle is about $r = (55 \pm 4)$ nm.

\begin{figure*}
\includegraphics[width=7in]{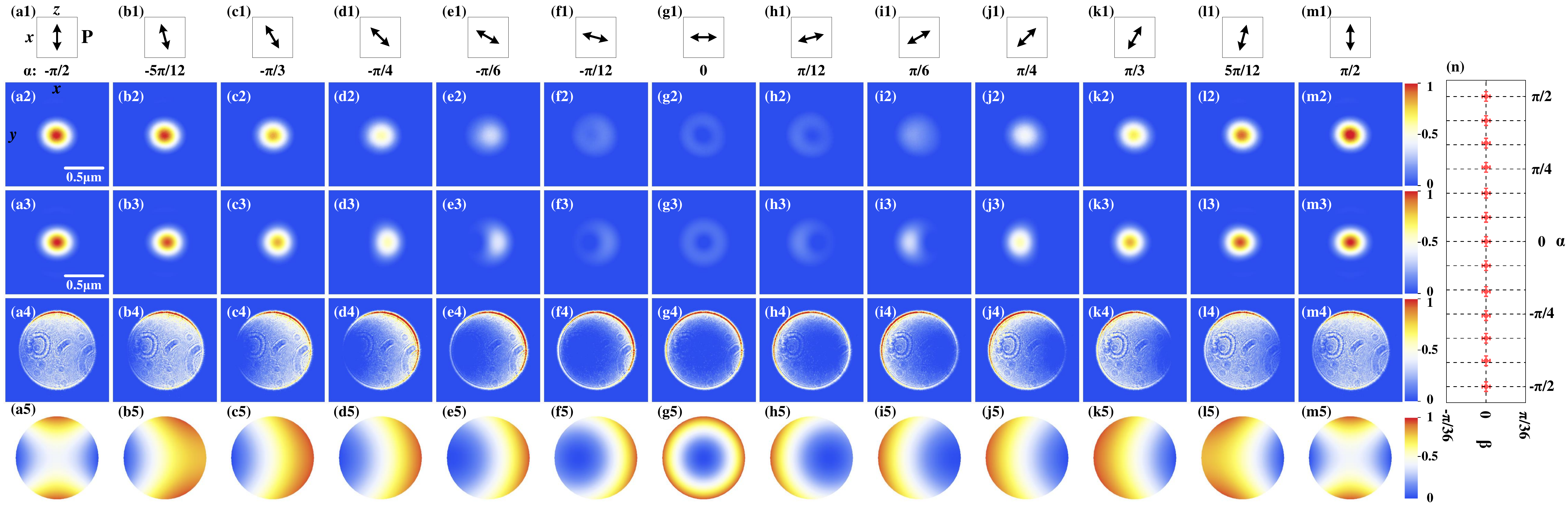}
\caption{The intensity distributions of the scattering light vs the linear polarization of the illuminating light in the image and Fourier planes for Case 1. (a1)-(m1) show the polarization of the illuminating light. (a2)-(m2) are the measured results in the image plane. (a3)-(m3) are the corresponding theoretical simulations in the image plane. (a4)-(m4) are the measured results  in Fourier space. (a5)-(m5) are the corresponding theoretical calculations in the Fourier plane. (n) The dipole orientations (dots) ($\alpha,\beta$) extracted from the measured k-space images compared to the polarization angles (intersections of dashed lines) of the incident light. A few specks in (a4)-(m4) is caused by some dusts on the microscope objective.}
\label{fig3}
\end{figure*}

We use a linearly polarized 532 nm laser beam (focused by a spherical lens $f = 175$ mm) to illuminate the optically trapped nanoparticle. The propagation direction of the illuminating beam is orthogonal to the optical axis of the microscope objective (the propagation direction of the trapping laser beam). The waist of the beam at the nanoparticle position is about 50 $\mu m$, which is much larger than the radius and displacement of the optically trapped nanoparticle. The 20 $\mu W$ power of this beam is weak enough to have any influence on the natural motion of the trapped nanoparticle. This illuminating light passes through a half-wave plate HW2 (before the vacuum chamber) for the adjustment of the input polarization. The light scattered by the nanoparticle is collected by the same microscope objective, and measured in the image and Fourier space on a CCD camera (Andor Zyla 4.2 sCMOS) by a 4f imaging system. Therefore, we obtain the image of the nanoparticle scattered light perpendicular to the propagation direction of the illumination beam. We use two detection schemes, case 1 with no PBS in the detection system (solid box and dashed box) is the polarization-independent detection scheme, and case 2 using half-wave plate and PBS combination (dotted box) is the polarization-dependent detection scheme. The case 2 of the detection scheme is used to study the polarization characteristics of the scattered light.

\textbf{Case 1 of the polarization-independent detection scheme:} First we discuss the images of the light scattered by the nanoparticle at different input linear polarizations of the illuminating light (532 nm) without any polarizing optical elements in the detection. We first adjust the half-wave plate HW3 to make the linear polarization along the $x$-axis, i.e. $\alpha=-\pi/2$, which corresponds to observing the 3D doughnut-shaped dipole scattering in the direction perpendicular to the axis of the doughnut. The scattering light distribution is measured in the image and the Fourier planes, respectively. The image profile of the scattered light for this polarization is circular and the intensity distribution exhibits a symmetric profile, as shown in Fig. 3(a2). From this intensity profile of the dipole scattering, we can not judge the dipole scattering orientations. In parallel, the scattering light pattern in the Fourier plane is measured as shown in Fig. 3(a4). This k-space profile of the dipole scattering can perfectly determine the dipole scattering orientations using the intensity distribution analysis. Then we change the polarization of the illuminating light in the anti-clockwise direction in the $xz$-plane and record the image of the scattered light for every $\pi/12$ radians of the polarization angle, as shown in Figs. 3(a2)-(m2) in the image plane and Figs. 3(a4)-(m4) in the Fourier plane. The intensity of the scattered light gradually becomes weaker as we scan the polarization from $-\pi/2$ to 0. The image profile of the scattered light changes into a doughnut hole for the input polarization of 0 (Figs. 3 (g2) and (g4)), and the intensity becomes the weakest. This corresponds to observing the 3D doughnut-shaped dipole scattering in the direction parallel to the axis of the doughnut. The doughnut hole lies in the center of the symmetrical intensity distribution for the illumination light polarization of 0, but it is tilted left and right for the polarization values of $-\pi/12$ and $\pi/12$ respectively (Figs. 3(f2), (f4), (h2), and (h4)). Figs. 3(a3)-(m3) and (a5)-(m5) are the theoretical simulations of the intensity distributions in the image and Fourier planes. Thus, we exactly determine the dipole orientations from the dipole scattering of the trapped nanoparticle by the scattering light pattern in the Fourier plane, as shown in Fig. 3(n).

\textbf{Case 2 of the polarization-dependent detection scheme:} Now we analyze the polarization distribution of the scattering light with the polarization components (half-wave plate and PBS). We first determine the polarization distribution of the scattered light (Fig. 4(a)) at the illumination light polarization of $\pi/2$. We put a half-wave plate HW3 and a PBS in front of the CCD, as shown in the dotted box in Fig. 2. By rotating the HW3, we extract different polarization components of the scattering light and record the images of the scattering light for every $\pi/6$ radians, as shown in Fig. 4(c). The intensity of the images gradually weakens from 0 to $\pi/2$ and increases back from $\pi/2$ to $\pi$. We can see that the intensity of the whole profile changes symmetrically with changing polarization, implying that the polarization distribution of the scattering light (Fig. 4(a)) exhibits a homogeneously linear polarization in space, as shown in Fig. 4(b).

\begin{figure}
\includegraphics[width=3in]{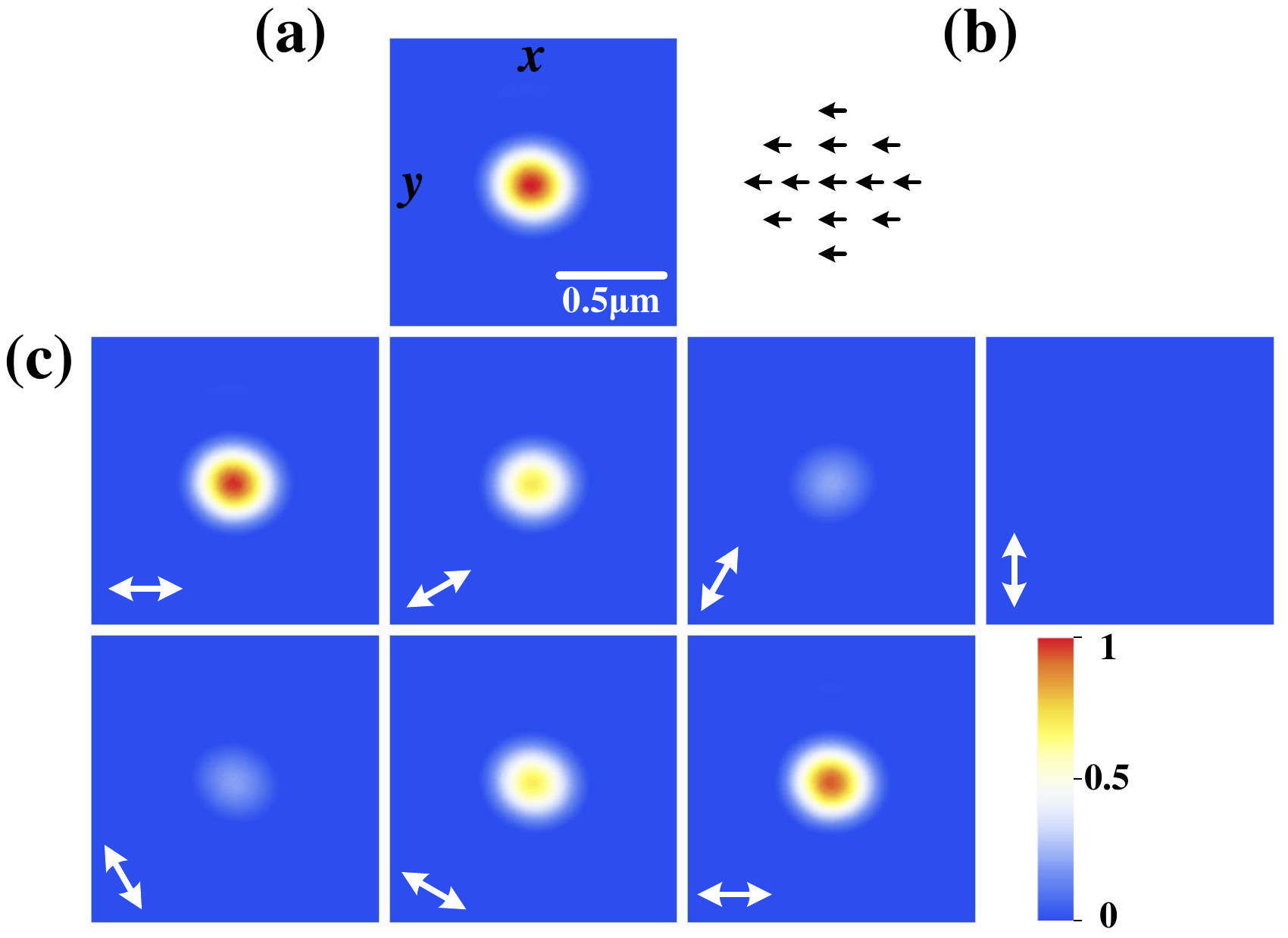}
\caption{The scattering light distribution in the image plane for Case 2 when $\alpha=\pi/2$. (a) shows the intensity distribution of the scattering light for Case 1 and (b) is the corresponding polarization distribution in the image plane. (c) The scattering distribution in the image plane using the polarization-dependent detection when the half wave plate HW3 is rotated. The arrows denote the detected polarization components.}
\label{fig:4}
\end{figure}

Next we determine the polarization distribution of the scattered light (Fig. 5(a)) at the illumination light polarization of $0$. When the polarization is aligned along horizontal direction, the scattering image splits into two parts, as shown in Fig. 5(c). The orientation of the two parts rotates following the polarization alignment of the scattering light while its intensity does not change. This shows that the polarization distribution of the scattering light (Fig. 5(a)) exhibits a vector light with radial polarization, as shown in Fig. 5(b).

\begin{figure}
\includegraphics[width=3in]{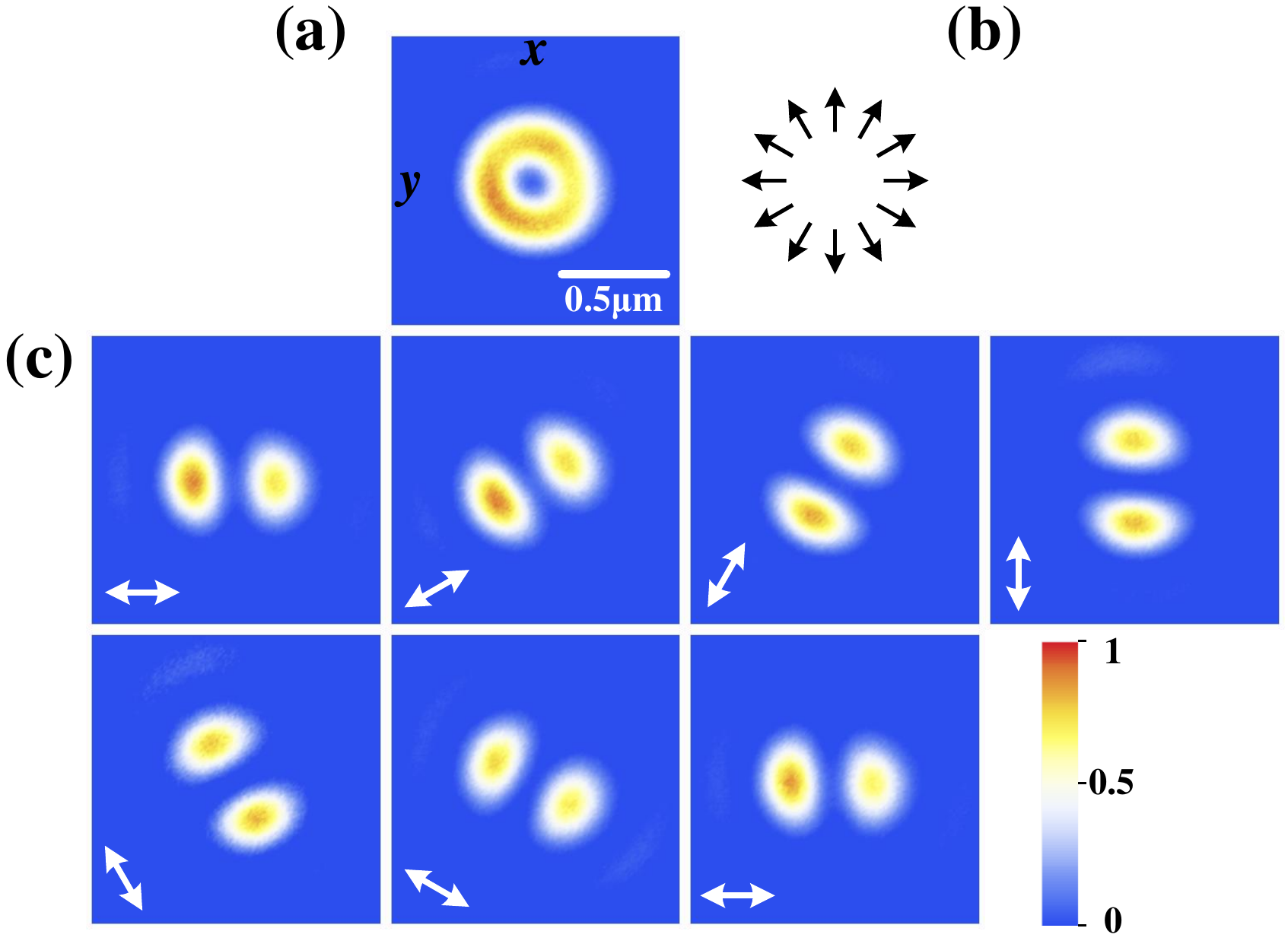}
\caption{The scattering light distribution in the image plane for Case 2 when $\alpha=0$. (a) shows the scattering light distribution for Case 1 and (b) is the corresponding polarization distribution in the image plane. (c) The scattering distribution in the image plane using the polarization-dependent detection scheme when the half wave plate HW3 is rotated. The arrows denote the detected polarization components.}
\label{fig:5}
\end{figure}

In conclusion, we have experimentally observed the dipole scattering of an optically levitated dielectric nanoparticle in vacuum by a strongly focused laser beam using a single high NA lens for the trapping of the particle and the detection of the scattering light. We have measured the intensity distribution in the image and the Fourier plane and the polarization distribution of the scattered light of the nanoparticle for the different linear polarizations of the illuminating light to determine the 3D dipole orientation. Although we only align the dipole axis along the meridian of the Bloch sphere, this work can be extended to align the dipole along any arbitrary direction by just changing the propagation direction of the illuminating light. Our work provides an environment free of particle-substrate interaction to image the dipole scattering pattern of a nanoparticle. This platform is useful in the future study of the scattering anisotropy with Kerker conditions for high-refractive index dielectric nanoparticles.

\begin{acknowledgments}
We thank Dr. Khan Sadiq Nawaz for the helpful discussion. This research is supported by the National Natural Science Foundation of China (Grants No. 12034011, No. 61975101), the Shanxi "1331 Project" Key Subjects Construction and the Xplorer Prize.
\end{acknowledgments}

\section*{data availability}
The data that support the findings of this study are available from the corresponding author upon reasonable request.

\nocite{*}
\bibliography{reference}

\begin{thebibliography}{49}%
\makeatletter
\providecommand \@ifxundefined [1]{%
 \@ifx{#1\undefined}
}%
\providecommand \@ifnum [1]{%
 \ifnum #1\expandafter \@firstoftwo
 \else \expandafter \@secondoftwo
 \fi
}%
\providecommand \@ifx [1]{%
 \ifx #1\expandafter \@firstoftwo
 \else \expandafter \@secondoftwo
 \fi
}%
\providecommand \natexlab [1]{#1}%
\providecommand \enquote  [1]{``#1''}%
\providecommand \bibnamefont  [1]{#1}%
\providecommand \bibfnamefont [1]{#1}%
\providecommand \citenamefont [1]{#1}%
\providecommand \href@noop [0]{\@secondoftwo}%
\providecommand \href [0]{\begingroup \@sanitize@url \@href}%
\providecommand \@href[1]{\@@startlink{#1}\@@href}%
\providecommand \@@href[1]{\endgroup#1\@@endlink}%
\providecommand \@sanitize@url [0]{\catcode `\\12\catcode `\$12\catcode
  `\&12\catcode `\#12\catcode `\^12\catcode `\_12\catcode `\%12\relax}%
\providecommand \@@startlink[1]{}%
\providecommand \@@endlink[0]{}%
\providecommand \url  [0]{\begingroup\@sanitize@url \@url }%
\providecommand \@url [1]{\endgroup\@href {#1}{\urlprefix }}%
\providecommand \urlprefix  [0]{URL }%
\providecommand \Eprint [0]{\href }%
\providecommand \doibase [0]{http://dx.doi.org/}%
\providecommand \selectlanguage [0]{\@gobble}%
\providecommand \bibinfo  [0]{\@secondoftwo}%
\providecommand \bibfield  [0]{\@secondoftwo}%
\providecommand \translation [1]{[#1]}%
\providecommand \BibitemOpen [0]{}%
\providecommand \bibitemStop [0]{}%
\providecommand \bibitemNoStop [0]{.\EOS\space}%
\providecommand \EOS [0]{\spacefactor3000\relax}%
\providecommand \BibitemShut  [1]{\csname bibitem#1\endcsname}%
\let\auto@bib@innerbib\@empty
\bibitem [{\citenamefont {Ashkin}(1970)}]{PhysRevLett.24.156}%
  \BibitemOpen
  \bibfield  {author} {\bibinfo {author} {\bibfnamefont {A.}~\bibnamefont
  {Ashkin}},\ }\href {\doibase 10.1103/PhysRevLett.24.156} {\bibfield
  {journal} {\bibinfo  {journal} {Phys. Rev. Lett.}\ }\textbf {\bibinfo
  {volume} {24}},\ \bibinfo {pages} {156} (\bibinfo {year} {1970})}\BibitemShut
  {NoStop}%
\bibitem [{\citenamefont {Ashkin}\ and\ \citenamefont
  {Dziedzic}(1976)}]{doi:10.1063/1.88748}%
  \BibitemOpen
  \bibfield  {author} {\bibinfo {author} {\bibfnamefont {A.}~\bibnamefont
  {Ashkin}}\ and\ \bibinfo {author} {\bibfnamefont {J.~M.}\ \bibnamefont
  {Dziedzic}},\ }\href {\doibase 10.1063/1.88748} {\bibfield  {journal}
  {\bibinfo  {journal} {Appl. Phys. Lett.}\ }\textbf {\bibinfo {volume} {28}},\
  \bibinfo {pages} {333} (\bibinfo {year} {1976})}\BibitemShut {NoStop}%
\bibitem [{\citenamefont {Yin}\ \emph {et~al.}(2013)\citenamefont {Yin},
  \citenamefont {Geraci},\ and\ \citenamefont
  {Li}}]{doi:10.1142/S0217979213300181}%
  \BibitemOpen
  \bibfield  {author} {\bibinfo {author} {\bibfnamefont {Z.-Q.}\ \bibnamefont
  {Yin}}, \bibinfo {author} {\bibfnamefont {A.~A.}\ \bibnamefont {Geraci}}, \
  and\ \bibinfo {author} {\bibfnamefont {T.}~\bibnamefont {Li}},\ }\href
  {\doibase 10.1142/S0217979213300181} {\bibfield  {journal} {\bibinfo
  {journal} {Int. J. Mod. Phys. B}\ }\textbf {\bibinfo {volume} {27}},\
  \bibinfo {pages} {1330018} (\bibinfo {year} {2013})}\BibitemShut {NoStop}%
\bibitem [{\citenamefont {Millen}\ \emph {et~al.}(2020)\citenamefont {Millen},
  \citenamefont {Monteiro}, \citenamefont {Pettit},\ and\ \citenamefont
  {Vamivakas}}]{Millen_2020}%
  \BibitemOpen
  \bibfield  {author} {\bibinfo {author} {\bibfnamefont {J.}~\bibnamefont
  {Millen}}, \bibinfo {author} {\bibfnamefont {T.~S.}\ \bibnamefont
  {Monteiro}}, \bibinfo {author} {\bibfnamefont {R.}~\bibnamefont {Pettit}}, \
  and\ \bibinfo {author} {\bibfnamefont {A.~N.}\ \bibnamefont {Vamivakas}},\
  }\href {\doibase 10.1088/1361-6633/ab6100} {\bibfield  {journal} {\bibinfo
  {journal} {Rep. Prog. Phys.}\ }\textbf {\bibinfo {volume} {83}},\ \bibinfo
  {pages} {026401} (\bibinfo {year} {2020})}\BibitemShut {NoStop}%
\bibitem [{\citenamefont {Geraci}\ \emph {et~al.}(2010)\citenamefont {Geraci},
  \citenamefont {Papp},\ and\ \citenamefont
  {Kitching}}]{PhysRevLett.105.101101}%
  \BibitemOpen
  \bibfield  {author} {\bibinfo {author} {\bibfnamefont {A.~A.}\ \bibnamefont
  {Geraci}}, \bibinfo {author} {\bibfnamefont {S.~B.}\ \bibnamefont {Papp}}, \
  and\ \bibinfo {author} {\bibfnamefont {J.}~\bibnamefont {Kitching}},\ }\href
  {\doibase 10.1103/PhysRevLett.105.101101} {\bibfield  {journal} {\bibinfo
  {journal} {Phys. Rev. Lett.}\ }\textbf {\bibinfo {volume} {105}},\ \bibinfo
  {pages} {101101} (\bibinfo {year} {2010})}\BibitemShut {NoStop}%
\bibitem [{\citenamefont {Winstone}\ \emph {et~al.}(2018)\citenamefont
  {Winstone}, \citenamefont {Bennett}, \citenamefont {Rademacher},
  \citenamefont {Rashid}, \citenamefont {Buhmann},\ and\ \citenamefont
  {Ulbricht}}]{PhysRevA.98.053831}%
  \BibitemOpen
  \bibfield  {author} {\bibinfo {author} {\bibfnamefont {G.}~\bibnamefont
  {Winstone}}, \bibinfo {author} {\bibfnamefont {R.}~\bibnamefont {Bennett}},
  \bibinfo {author} {\bibfnamefont {M.}~\bibnamefont {Rademacher}}, \bibinfo
  {author} {\bibfnamefont {M.}~\bibnamefont {Rashid}}, \bibinfo {author}
  {\bibfnamefont {S.}~\bibnamefont {Buhmann}}, \ and\ \bibinfo {author}
  {\bibfnamefont {H.}~\bibnamefont {Ulbricht}},\ }\href {\doibase
  10.1103/PhysRevA.98.053831} {\bibfield  {journal} {\bibinfo  {journal} {Phys.
  Rev. A}\ }\textbf {\bibinfo {volume} {98}},\ \bibinfo {pages} {053831}
  (\bibinfo {year} {2018})}\BibitemShut {NoStop}%
\bibitem [{\citenamefont {Hebestreit}\ \emph {et~al.}(2018)\citenamefont
  {Hebestreit}, \citenamefont {Frimmer}, \citenamefont {Reimann},\ and\
  \citenamefont {Novotny}}]{PhysRevLett.121.063602}%
  \BibitemOpen
  \bibfield  {author} {\bibinfo {author} {\bibfnamefont {E.}~\bibnamefont
  {Hebestreit}}, \bibinfo {author} {\bibfnamefont {M.}~\bibnamefont {Frimmer}},
  \bibinfo {author} {\bibfnamefont {R.}~\bibnamefont {Reimann}}, \ and\
  \bibinfo {author} {\bibfnamefont {L.}~\bibnamefont {Novotny}},\ }\href
  {\doibase 10.1103/PhysRevLett.121.063602} {\bibfield  {journal} {\bibinfo
  {journal} {Phys. Rev. Lett.}\ }\textbf {\bibinfo {volume} {121}},\ \bibinfo
  {pages} {063602} (\bibinfo {year} {2018})}\BibitemShut {NoStop}%
\bibitem [{\citenamefont {Timberlake}\ \emph {et~al.}(2019)\citenamefont
  {Timberlake}, \citenamefont {Toroš}, \citenamefont {Hempston}, \citenamefont
  {Winstone}, \citenamefont {Rashid},\ and\ \citenamefont
  {Ulbricht}}]{doi:10.1063/1.5081045}%
  \BibitemOpen
  \bibfield  {author} {\bibinfo {author} {\bibfnamefont {C.}~\bibnamefont
  {Timberlake}}, \bibinfo {author} {\bibfnamefont {M.}~\bibnamefont {Toroš}},
  \bibinfo {author} {\bibfnamefont {D.}~\bibnamefont {Hempston}}, \bibinfo
  {author} {\bibfnamefont {G.}~\bibnamefont {Winstone}}, \bibinfo {author}
  {\bibfnamefont {M.}~\bibnamefont {Rashid}}, \ and\ \bibinfo {author}
  {\bibfnamefont {H.}~\bibnamefont {Ulbricht}},\ }\href {\doibase
  10.1063/1.5081045} {\bibfield  {journal} {\bibinfo  {journal} {Appl. Phys.
  Lett.}\ }\textbf {\bibinfo {volume} {114}},\ \bibinfo {pages} {023104}
  (\bibinfo {year} {2019})}\BibitemShut {NoStop}%
\bibitem [{\citenamefont {Li}\ \emph {et~al.}(2011)\citenamefont {Li},
  \citenamefont {Kheifets},\ and\ \citenamefont {Raizen}}]{NatPhys.7.527}%
  \BibitemOpen
  \bibfield  {author} {\bibinfo {author} {\bibfnamefont {T.}~\bibnamefont
  {Li}}, \bibinfo {author} {\bibfnamefont {S.}~\bibnamefont {Kheifets}}, \ and\
  \bibinfo {author} {\bibfnamefont {M.~G.}\ \bibnamefont {Raizen}},\ }\href
  {\doibase 10.1038/nphys1952} {\bibfield  {journal} {\bibinfo  {journal} {Nat.
  Phys.}\ }\textbf {\bibinfo {volume} {7}},\ \bibinfo {pages} {527} (\bibinfo
  {year} {2011})}\BibitemShut {NoStop}%
\bibitem [{\citenamefont {Gieseler}\ \emph {et~al.}(2012)\citenamefont
  {Gieseler}, \citenamefont {Deutsch}, \citenamefont {Quidant},\ and\
  \citenamefont {Novotny}}]{PhysRevLett.109.103603}%
  \BibitemOpen
  \bibfield  {author} {\bibinfo {author} {\bibfnamefont {J.}~\bibnamefont
  {Gieseler}}, \bibinfo {author} {\bibfnamefont {B.}~\bibnamefont {Deutsch}},
  \bibinfo {author} {\bibfnamefont {R.}~\bibnamefont {Quidant}}, \ and\
  \bibinfo {author} {\bibfnamefont {L.}~\bibnamefont {Novotny}},\ }\href
  {\doibase 10.1103/PhysRevLett.109.103603} {\bibfield  {journal} {\bibinfo
  {journal} {Phys. Rev. Lett.}\ }\textbf {\bibinfo {volume} {109}},\ \bibinfo
  {pages} {103603} (\bibinfo {year} {2012})}\BibitemShut {NoStop}%
\bibitem [{\citenamefont {Vovrosh}\ \emph {et~al.}(2017)\citenamefont
  {Vovrosh}, \citenamefont {Rashid}, \citenamefont {Hempston}, \citenamefont
  {Bateman}, \citenamefont {Paternostro},\ and\ \citenamefont
  {Ulbricht}}]{Vovrosh:s}%
  \BibitemOpen
  \bibfield  {author} {\bibinfo {author} {\bibfnamefont {J.}~\bibnamefont
  {Vovrosh}}, \bibinfo {author} {\bibfnamefont {M.}~\bibnamefont {Rashid}},
  \bibinfo {author} {\bibfnamefont {D.}~\bibnamefont {Hempston}}, \bibinfo
  {author} {\bibfnamefont {J.}~\bibnamefont {Bateman}}, \bibinfo {author}
  {\bibfnamefont {M.}~\bibnamefont {Paternostro}}, \ and\ \bibinfo {author}
  {\bibfnamefont {H.}~\bibnamefont {Ulbricht}},\ }\href {\doibase
  10.1364/JOSAB.34.001421} {\bibfield  {journal} {\bibinfo  {journal} {J. Opt.
  Soc. Am. B}\ }\textbf {\bibinfo {volume} {34}},\ \bibinfo {pages} {1421}
  (\bibinfo {year} {2017})}\BibitemShut {NoStop}%
\bibitem [{\citenamefont {Zheng}\ \emph {et~al.}(2019)\citenamefont {Zheng},
  \citenamefont {Guo},\ and\ \citenamefont {Sun}}]{doi:10.1063/1.5099284}%
  \BibitemOpen
  \bibfield  {author} {\bibinfo {author} {\bibfnamefont {Y.}~\bibnamefont
  {Zheng}}, \bibinfo {author} {\bibfnamefont {G.}~\bibnamefont {Guo}}, \ and\
  \bibinfo {author} {\bibfnamefont {F.}~\bibnamefont {Sun}},\ }\href {\doibase
  10.1063/1.5099284} {\bibfield  {journal} {\bibinfo  {journal} {Appl. Phys.
  Lett.}\ }\textbf {\bibinfo {volume} {115}},\ \bibinfo {pages} {101105}
  (\bibinfo {year} {2019})}\BibitemShut {NoStop}%
\bibitem [{\citenamefont {Iwasaki}\ \emph {et~al.}(2019)\citenamefont
  {Iwasaki}, \citenamefont {Yotsuya}, \citenamefont {Naruki}, \citenamefont
  {Matsuda}, \citenamefont {Yoneda},\ and\ \citenamefont
  {Aikawa}}]{PhysRevA.99.051401}%
  \BibitemOpen
  \bibfield  {author} {\bibinfo {author} {\bibfnamefont {M.}~\bibnamefont
  {Iwasaki}}, \bibinfo {author} {\bibfnamefont {T.}~\bibnamefont {Yotsuya}},
  \bibinfo {author} {\bibfnamefont {T.}~\bibnamefont {Naruki}}, \bibinfo
  {author} {\bibfnamefont {Y.}~\bibnamefont {Matsuda}}, \bibinfo {author}
  {\bibfnamefont {M.}~\bibnamefont {Yoneda}}, \ and\ \bibinfo {author}
  {\bibfnamefont {K.}~\bibnamefont {Aikawa}},\ }\href {\doibase
  10.1103/PhysRevA.99.051401} {\bibfield  {journal} {\bibinfo  {journal} {Phys.
  Rev. A}\ }\textbf {\bibinfo {volume} {99}},\ \bibinfo {pages} {051401}
  (\bibinfo {year} {2019})}\BibitemShut {NoStop}%
\bibitem [{\citenamefont {Tebbenjohanns}\ \emph {et~al.}(2019)\citenamefont
  {Tebbenjohanns}, \citenamefont {Frimmer}, \citenamefont {Militaru},
  \citenamefont {Jain},\ and\ \citenamefont
  {Novotny}}]{PhysRevLett.122.223601}%
  \BibitemOpen
  \bibfield  {author} {\bibinfo {author} {\bibfnamefont {F.}~\bibnamefont
  {Tebbenjohanns}}, \bibinfo {author} {\bibfnamefont {M.}~\bibnamefont
  {Frimmer}}, \bibinfo {author} {\bibfnamefont {A.}~\bibnamefont {Militaru}},
  \bibinfo {author} {\bibfnamefont {V.}~\bibnamefont {Jain}}, \ and\ \bibinfo
  {author} {\bibfnamefont {L.}~\bibnamefont {Novotny}},\ }\href {\doibase
  10.1103/PhysRevLett.122.223601} {\bibfield  {journal} {\bibinfo  {journal}
  {Phys. Rev. Lett.}\ }\textbf {\bibinfo {volume} {122}},\ \bibinfo {pages}
  {223601} (\bibinfo {year} {2019})}\BibitemShut {NoStop}%
\bibitem [{\citenamefont {Conangla}\ \emph {et~al.}(2019)\citenamefont
  {Conangla}, \citenamefont {Ricci}, \citenamefont {Cuairan}, \citenamefont
  {Schell}, \citenamefont {Meyer},\ and\ \citenamefont
  {Quidant}}]{PhysRevLett.122.223602}%
  \BibitemOpen
  \bibfield  {author} {\bibinfo {author} {\bibfnamefont {G.~P.}\ \bibnamefont
  {Conangla}}, \bibinfo {author} {\bibfnamefont {F.}~\bibnamefont {Ricci}},
  \bibinfo {author} {\bibfnamefont {M.~T.}\ \bibnamefont {Cuairan}}, \bibinfo
  {author} {\bibfnamefont {A.~W.}\ \bibnamefont {Schell}}, \bibinfo {author}
  {\bibfnamefont {N.}~\bibnamefont {Meyer}}, \ and\ \bibinfo {author}
  {\bibfnamefont {R.}~\bibnamefont {Quidant}},\ }\href {\doibase
  10.1103/PhysRevLett.122.223602} {\bibfield  {journal} {\bibinfo  {journal}
  {Phys. Rev. Lett.}\ }\textbf {\bibinfo {volume} {122}},\ \bibinfo {pages}
  {223602} (\bibinfo {year} {2019})}\BibitemShut {NoStop}%
\bibitem [{\citenamefont {Windey}\ \emph {et~al.}(2019)\citenamefont {Windey},
  \citenamefont {Gonzalez-Ballestero}, \citenamefont {Maurer}, \citenamefont
  {Novotny}, \citenamefont {Romero-Isart},\ and\ \citenamefont
  {Reimann}}]{PhysRevLett.122.123601}%
  \BibitemOpen
  \bibfield  {author} {\bibinfo {author} {\bibfnamefont {D.}~\bibnamefont
  {Windey}}, \bibinfo {author} {\bibfnamefont {C.}~\bibnamefont
  {Gonzalez-Ballestero}}, \bibinfo {author} {\bibfnamefont {P.}~\bibnamefont
  {Maurer}}, \bibinfo {author} {\bibfnamefont {L.}~\bibnamefont {Novotny}},
  \bibinfo {author} {\bibfnamefont {O.}~\bibnamefont {Romero-Isart}}, \ and\
  \bibinfo {author} {\bibfnamefont {R.}~\bibnamefont {Reimann}},\ }\href
  {\doibase 10.1103/PhysRevLett.122.123601} {\bibfield  {journal} {\bibinfo
  {journal} {Phys. Rev. Lett.}\ }\textbf {\bibinfo {volume} {122}},\ \bibinfo
  {pages} {123601} (\bibinfo {year} {2019})}\BibitemShut {NoStop}%
\bibitem [{\citenamefont {Deli{\'c}}\ \emph {et~al.}(2019)\citenamefont
  {Deli{\'c}}, \citenamefont {Reisenbauer}, \citenamefont {Grass},
  \citenamefont {Kiesel}, \citenamefont {Vuleti{\'c}},\ and\ \citenamefont
  {Aspelmeyer}}]{PhysRevLett.122.123602}%
  \BibitemOpen
  \bibfield  {author} {\bibinfo {author} {\bibfnamefont {U.}~\bibnamefont
  {Deli{\'c}}}, \bibinfo {author} {\bibfnamefont {M.}~\bibnamefont
  {Reisenbauer}}, \bibinfo {author} {\bibfnamefont {D.}~\bibnamefont {Grass}},
  \bibinfo {author} {\bibfnamefont {N.}~\bibnamefont {Kiesel}}, \bibinfo
  {author} {\bibfnamefont {V.}~\bibnamefont {Vuleti{\'c}}}, \ and\ \bibinfo
  {author} {\bibfnamefont {M.}~\bibnamefont {Aspelmeyer}},\ }\href {\doibase
  10.1103/PhysRevLett.122.123602} {\bibfield  {journal} {\bibinfo  {journal}
  {Phys. Rev. Lett.}\ }\textbf {\bibinfo {volume} {122}},\ \bibinfo {pages}
  {123602} (\bibinfo {year} {2019})}\BibitemShut {NoStop}%
\bibitem [{\citenamefont {Deli{\'c}}\ \emph {et~al.}(2020)\citenamefont
  {Deli{\'c}}, \citenamefont {Reisenbauer}, \citenamefont {Dare}, \citenamefont
  {Grass}, \citenamefont {Vuleti{\'c}}, \citenamefont {Kiesel},\ and\
  \citenamefont {Aspelmeyer}}]{Deli892}%
  \BibitemOpen
  \bibfield  {author} {\bibinfo {author} {\bibfnamefont {U.}~\bibnamefont
  {Deli{\'c}}}, \bibinfo {author} {\bibfnamefont {M.}~\bibnamefont
  {Reisenbauer}}, \bibinfo {author} {\bibfnamefont {K.}~\bibnamefont {Dare}},
  \bibinfo {author} {\bibfnamefont {D.}~\bibnamefont {Grass}}, \bibinfo
  {author} {\bibfnamefont {V.}~\bibnamefont {Vuleti{\'c}}}, \bibinfo {author}
  {\bibfnamefont {N.}~\bibnamefont {Kiesel}}, \ and\ \bibinfo {author}
  {\bibfnamefont {M.}~\bibnamefont {Aspelmeyer}},\ }\href {\doibase
  10.1126/science.aba3993} {\bibfield  {journal} {\bibinfo  {journal}
  {Science}\ }\textbf {\bibinfo {volume} {367}},\ \bibinfo {pages} {892}
  (\bibinfo {year} {2020})}\BibitemShut {NoStop}%
\bibitem [{\citenamefont {Romero-Isart}\ \emph {et~al.}(2010)\citenamefont
  {Romero-Isart}, \citenamefont {Juan}, \citenamefont {Quidant},\ and\
  \citenamefont {Cirac}}]{Romero_Isart_2010}%
  \BibitemOpen
  \bibfield  {author} {\bibinfo {author} {\bibfnamefont {O.}~\bibnamefont
  {Romero-Isart}}, \bibinfo {author} {\bibfnamefont {M.~L.}\ \bibnamefont
  {Juan}}, \bibinfo {author} {\bibfnamefont {R.}~\bibnamefont {Quidant}}, \
  and\ \bibinfo {author} {\bibfnamefont {J.~I.}\ \bibnamefont {Cirac}},\ }\href
  {\doibase 10.1088/1367-2630/12/3/033015} {\bibfield  {journal} {\bibinfo
  {journal} {New J. Phys.}\ }\textbf {\bibinfo {volume} {12}},\ \bibinfo
  {pages} {033015} (\bibinfo {year} {2010})}\BibitemShut {NoStop}%
\bibitem [{\citenamefont {Romero-Isart}\ \emph {et~al.}(2011)\citenamefont
  {Romero-Isart}, \citenamefont {Pflanzer}, \citenamefont {Blaser},
  \citenamefont {Kaltenbaek}, \citenamefont {Kiesel}, \citenamefont
  {Aspelmeyer},\ and\ \citenamefont {Cirac}}]{Romero-Isart11}%
  \BibitemOpen
  \bibfield  {author} {\bibinfo {author} {\bibfnamefont {O.}~\bibnamefont
  {Romero-Isart}}, \bibinfo {author} {\bibfnamefont {A.~C.}\ \bibnamefont
  {Pflanzer}}, \bibinfo {author} {\bibfnamefont {F.}~\bibnamefont {Blaser}},
  \bibinfo {author} {\bibfnamefont {R.}~\bibnamefont {Kaltenbaek}}, \bibinfo
  {author} {\bibfnamefont {N.}~\bibnamefont {Kiesel}}, \bibinfo {author}
  {\bibfnamefont {M.}~\bibnamefont {Aspelmeyer}}, \ and\ \bibinfo {author}
  {\bibfnamefont {J.~I.}\ \bibnamefont {Cirac}},\ }\href {\doibase
  10.1103/PhysRevLett.107.020405} {\bibfield  {journal} {\bibinfo  {journal}
  {Phys. Rev. Lett.}\ }\textbf {\bibinfo {volume} {107}},\ \bibinfo {pages}
  {020405} (\bibinfo {year} {2011})}\BibitemShut {NoStop}%
\bibitem [{\citenamefont {Wan}\ \emph {et~al.}(2016)\citenamefont {Wan},
  \citenamefont {Scala}, \citenamefont {Morley}, \citenamefont {Rahman},
  \citenamefont {Ulbricht}, \citenamefont {Bateman}, \citenamefont {Barker},
  \citenamefont {Bose},\ and\ \citenamefont {Kim}}]{Wan16}%
  \BibitemOpen
  \bibfield  {author} {\bibinfo {author} {\bibfnamefont {C.}~\bibnamefont
  {Wan}}, \bibinfo {author} {\bibfnamefont {M.}~\bibnamefont {Scala}}, \bibinfo
  {author} {\bibfnamefont {G.~W.}\ \bibnamefont {Morley}}, \bibinfo {author}
  {\bibfnamefont {A.~A.}\ \bibnamefont {Rahman}}, \bibinfo {author}
  {\bibfnamefont {H.}~\bibnamefont {Ulbricht}}, \bibinfo {author}
  {\bibfnamefont {J.}~\bibnamefont {Bateman}}, \bibinfo {author} {\bibfnamefont
  {P.~F.}\ \bibnamefont {Barker}}, \bibinfo {author} {\bibfnamefont
  {S.}~\bibnamefont {Bose}}, \ and\ \bibinfo {author} {\bibfnamefont {M.~S.}\
  \bibnamefont {Kim}},\ }\href {\doibase 10.1103/PhysRevLett.117.143003}
  {\bibfield  {journal} {\bibinfo  {journal} {Phys. Rev. Lett.}\ }\textbf
  {\bibinfo {volume} {117}},\ \bibinfo {pages} {143003} (\bibinfo {year}
  {2016})}\BibitemShut {NoStop}%
\bibitem [{\citenamefont {Ahn}\ \emph {et~al.}(2018)\citenamefont {Ahn},
  \citenamefont {Xu}, \citenamefont {Bang}, \citenamefont {Deng}, \citenamefont
  {Hoang}, \citenamefont {Han}, \citenamefont {Ma},\ and\ \citenamefont
  {Li}}]{PhysRevLett.121.033603}%
  \BibitemOpen
  \bibfield  {author} {\bibinfo {author} {\bibfnamefont {J.}~\bibnamefont
  {Ahn}}, \bibinfo {author} {\bibfnamefont {Z.}~\bibnamefont {Xu}}, \bibinfo
  {author} {\bibfnamefont {J.}~\bibnamefont {Bang}}, \bibinfo {author}
  {\bibfnamefont {Y.-H.}\ \bibnamefont {Deng}}, \bibinfo {author}
  {\bibfnamefont {T.~M.}\ \bibnamefont {Hoang}}, \bibinfo {author}
  {\bibfnamefont {Q.}~\bibnamefont {Han}}, \bibinfo {author} {\bibfnamefont
  {R.-M.}\ \bibnamefont {Ma}}, \ and\ \bibinfo {author} {\bibfnamefont
  {T.}~\bibnamefont {Li}},\ }\href {\doibase 10.1103/PhysRevLett.121.033603}
  {\bibfield  {journal} {\bibinfo  {journal} {Phys. Rev. Lett.}\ }\textbf
  {\bibinfo {volume} {121}},\ \bibinfo {pages} {033603} (\bibinfo {year}
  {2018})}\BibitemShut {NoStop}%
\bibitem [{\citenamefont {Xu}\ and\ \citenamefont
  {Li}(2017)}]{PhysRevA.96.033843}%
  \BibitemOpen
  \bibfield  {author} {\bibinfo {author} {\bibfnamefont {Z.}~\bibnamefont
  {Xu}}\ and\ \bibinfo {author} {\bibfnamefont {T.}~\bibnamefont {Li}},\ }\href
  {\doibase 10.1103/PhysRevA.96.033843} {\bibfield  {journal} {\bibinfo
  {journal} {Phys. Rev. A}\ }\textbf {\bibinfo {volume} {96}},\ \bibinfo
  {pages} {033843} (\bibinfo {year} {2017})}\BibitemShut {NoStop}%
\bibitem [{\citenamefont {Reimann}\ \emph {et~al.}(2018)\citenamefont
  {Reimann}, \citenamefont {Doderer}, \citenamefont {Hebestreit}, \citenamefont
  {Diehl}, \citenamefont {Frimmer}, \citenamefont {Windey}, \citenamefont
  {Tebbenjohanns},\ and\ \citenamefont {Novotny}}]{PhysRevLett.121.033602}%
  \BibitemOpen
  \bibfield  {author} {\bibinfo {author} {\bibfnamefont {R.}~\bibnamefont
  {Reimann}}, \bibinfo {author} {\bibfnamefont {M.}~\bibnamefont {Doderer}},
  \bibinfo {author} {\bibfnamefont {E.}~\bibnamefont {Hebestreit}}, \bibinfo
  {author} {\bibfnamefont {R.}~\bibnamefont {Diehl}}, \bibinfo {author}
  {\bibfnamefont {M.}~\bibnamefont {Frimmer}}, \bibinfo {author} {\bibfnamefont
  {D.}~\bibnamefont {Windey}}, \bibinfo {author} {\bibfnamefont
  {F.}~\bibnamefont {Tebbenjohanns}}, \ and\ \bibinfo {author} {\bibfnamefont
  {L.}~\bibnamefont {Novotny}},\ }\href {\doibase
  10.1103/PhysRevLett.121.033602} {\bibfield  {journal} {\bibinfo  {journal}
  {Phys. Rev. Lett.}\ }\textbf {\bibinfo {volume} {121}},\ \bibinfo {pages}
  {033602} (\bibinfo {year} {2018})}\BibitemShut {NoStop}%
\bibitem [{\citenamefont {Ahn}\ \emph {et~al.}(2020)\citenamefont {Ahn},
  \citenamefont {Xu}, \citenamefont {Bang}, \citenamefont {Ju}, \citenamefont
  {Gao},\ and\ \citenamefont {Li}}]{Li2020}%
  \BibitemOpen
  \bibfield  {author} {\bibinfo {author} {\bibfnamefont {J.}~\bibnamefont
  {Ahn}}, \bibinfo {author} {\bibfnamefont {Z.}~\bibnamefont {Xu}}, \bibinfo
  {author} {\bibfnamefont {J.}~\bibnamefont {Bang}}, \bibinfo {author}
  {\bibfnamefont {P.}~\bibnamefont {Ju}}, \bibinfo {author} {\bibfnamefont
  {X.}~\bibnamefont {Gao}}, \ and\ \bibinfo {author} {\bibfnamefont
  {T.}~\bibnamefont {Li}},\ }\href {\doibase 10.1038/s41565-019-0605-9}
  {\bibfield  {journal} {\bibinfo  {journal} {Nat. Nanotechnol.}\ }\textbf
  {\bibinfo {volume} {15}},\ \bibinfo {pages} {89} (\bibinfo {year}
  {2020})}\BibitemShut {NoStop}%
\bibitem [{\citenamefont {Jin}\ \emph {et~al.}(2021)\citenamefont {Jin},
  \citenamefont {Yan}, \citenamefont {Rahman}, \citenamefont {Li},
  \citenamefont {Yu},\ and\ \citenamefont {Zhang}}]{jin20206}%
  \BibitemOpen
  \bibfield  {author} {\bibinfo {author} {\bibfnamefont {Y.}~\bibnamefont
  {Jin}}, \bibinfo {author} {\bibfnamefont {J.}~\bibnamefont {Yan}}, \bibinfo
  {author} {\bibfnamefont {S.~J.}\ \bibnamefont {Rahman}}, \bibinfo {author}
  {\bibfnamefont {J.}~\bibnamefont {Li}}, \bibinfo {author} {\bibfnamefont
  {X.}~\bibnamefont {Yu}}, \ and\ \bibinfo {author} {\bibfnamefont
  {J.}~\bibnamefont {Zhang}},\ }\href {\doibase 10.1364/PRJ.422975} {\bibfield
  {journal} {\bibinfo  {journal} {Photon. Res.}\ }\textbf {\bibinfo {volume}
  {9}},\ \bibinfo {pages} {1344} (\bibinfo {year} {2021})}\BibitemShut
  {NoStop}%
\bibitem [{\citenamefont {Moskovits}(1985)}]{RevModPhys.57.783}%
  \BibitemOpen
  \bibfield  {author} {\bibinfo {author} {\bibfnamefont {M.}~\bibnamefont
  {Moskovits}},\ }\href {\doibase 10.1103/RevModPhys.57.783} {\bibfield
  {journal} {\bibinfo  {journal} {Rev. Mod. Phys.}\ }\textbf {\bibinfo {volume}
  {57}},\ \bibinfo {pages} {783} (\bibinfo {year} {1985})}\BibitemShut
  {NoStop}%
\bibitem [{\citenamefont {Kurilovich}\ \emph {et~al.}(2016)\citenamefont
  {Kurilovich}, \citenamefont {Klein}, \citenamefont {Torretti}, \citenamefont
  {Lassise}, \citenamefont {Hoekstra}, \citenamefont {Ubachs}, \citenamefont
  {Gelderblom},\ and\ \citenamefont {Versolato}}]{PhysRevApplied.6.014018}%
  \BibitemOpen
  \bibfield  {author} {\bibinfo {author} {\bibfnamefont {D.}~\bibnamefont
  {Kurilovich}}, \bibinfo {author} {\bibfnamefont {A.~L.}\ \bibnamefont
  {Klein}}, \bibinfo {author} {\bibfnamefont {F.}~\bibnamefont {Torretti}},
  \bibinfo {author} {\bibfnamefont {A.}~\bibnamefont {Lassise}}, \bibinfo
  {author} {\bibfnamefont {R.}~\bibnamefont {Hoekstra}}, \bibinfo {author}
  {\bibfnamefont {W.}~\bibnamefont {Ubachs}}, \bibinfo {author} {\bibfnamefont
  {H.}~\bibnamefont {Gelderblom}}, \ and\ \bibinfo {author} {\bibfnamefont
  {O.~O.}\ \bibnamefont {Versolato}},\ }\href {\doibase
  10.1103/PhysRevApplied.6.014018} {\bibfield  {journal} {\bibinfo  {journal}
  {Phys. Rev. Applied}\ }\textbf {\bibinfo {volume} {6}},\ \bibinfo {pages}
  {014018} (\bibinfo {year} {2016})}\BibitemShut {NoStop}%
\bibitem [{\citenamefont {Kurilovich}\ \emph {et~al.}(2018)\citenamefont
  {Kurilovich}, \citenamefont {Pinto}, \citenamefont {Torretti}, \citenamefont
  {Schupp}, \citenamefont {Scheers}, \citenamefont {Stodolna}, \citenamefont
  {Gelderblom}, \citenamefont {Eikema}, \citenamefont {Witte}, \citenamefont
  {Ubachs}, \citenamefont {Hoekstra},\ and\ \citenamefont
  {Versolato}}]{PhysRevApplied.10.054005}%
  \BibitemOpen
  \bibfield  {author} {\bibinfo {author} {\bibfnamefont {D.}~\bibnamefont
  {Kurilovich}}, \bibinfo {author} {\bibfnamefont {T.~d.~F.}\ \bibnamefont
  {Pinto}}, \bibinfo {author} {\bibfnamefont {F.}~\bibnamefont {Torretti}},
  \bibinfo {author} {\bibfnamefont {R.}~\bibnamefont {Schupp}}, \bibinfo
  {author} {\bibfnamefont {J.}~\bibnamefont {Scheers}}, \bibinfo {author}
  {\bibfnamefont {A.~S.}\ \bibnamefont {Stodolna}}, \bibinfo {author}
  {\bibfnamefont {H.}~\bibnamefont {Gelderblom}}, \bibinfo {author}
  {\bibfnamefont {K.~S.}\ \bibnamefont {Eikema}}, \bibinfo {author}
  {\bibfnamefont {S.}~\bibnamefont {Witte}}, \bibinfo {author} {\bibfnamefont
  {W.}~\bibnamefont {Ubachs}}, \bibinfo {author} {\bibfnamefont
  {R.}~\bibnamefont {Hoekstra}}, \ and\ \bibinfo {author} {\bibfnamefont
  {O.~O.}\ \bibnamefont {Versolato}},\ }\href {\doibase
  10.1103/PhysRevApplied.10.054005} {\bibfield  {journal} {\bibinfo  {journal}
  {Phys. Rev. Applied}\ }\textbf {\bibinfo {volume} {10}},\ \bibinfo {pages}
  {054005} (\bibinfo {year} {2018})}\BibitemShut {NoStop}%
\bibitem [{\citenamefont {Schupp}\ \emph {et~al.}(2019)\citenamefont {Schupp},
  \citenamefont {Torretti}, \citenamefont {Meijer}, \citenamefont {Bayraktar},
  \citenamefont {Scheers}, \citenamefont {Kurilovich}, \citenamefont {Bayerle},
  \citenamefont {Eikema}, \citenamefont {Witte}, \citenamefont {Ubachs},
  \citenamefont {Hoekstra},\ and\ \citenamefont
  {Versolato}}]{PhysRevApplied.12.014010}%
  \BibitemOpen
  \bibfield  {author} {\bibinfo {author} {\bibfnamefont {R.}~\bibnamefont
  {Schupp}}, \bibinfo {author} {\bibfnamefont {F.}~\bibnamefont {Torretti}},
  \bibinfo {author} {\bibfnamefont {R.}~\bibnamefont {Meijer}}, \bibinfo
  {author} {\bibfnamefont {M.}~\bibnamefont {Bayraktar}}, \bibinfo {author}
  {\bibfnamefont {J.}~\bibnamefont {Scheers}}, \bibinfo {author} {\bibfnamefont
  {D.}~\bibnamefont {Kurilovich}}, \bibinfo {author} {\bibfnamefont
  {A.}~\bibnamefont {Bayerle}}, \bibinfo {author} {\bibfnamefont
  {K.}~\bibnamefont {Eikema}}, \bibinfo {author} {\bibfnamefont
  {S.}~\bibnamefont {Witte}}, \bibinfo {author} {\bibfnamefont
  {W.}~\bibnamefont {Ubachs}}, \bibinfo {author} {\bibfnamefont
  {R.}~\bibnamefont {Hoekstra}}, \ and\ \bibinfo {author} {\bibfnamefont
  {O.}~\bibnamefont {Versolato}},\ }\href {\doibase
  10.1103/PhysRevApplied.12.014010} {\bibfield  {journal} {\bibinfo  {journal}
  {Phys. Rev. Applied}\ }\textbf {\bibinfo {volume} {12}},\ \bibinfo {pages}
  {014010} (\bibinfo {year} {2019})}\BibitemShut {NoStop}%
\bibitem [{\citenamefont {Fu}\ \emph {et~al.}(2013)\citenamefont {Fu},
  \citenamefont {Kuznetsov}, \citenamefont {Miroshnichenko}, \citenamefont
  {Yu},\ and\ \citenamefont {Luk’yanchuk}}]{ncomms2538}%
  \BibitemOpen
  \bibfield  {author} {\bibinfo {author} {\bibfnamefont {Y.~H.}\ \bibnamefont
  {Fu}}, \bibinfo {author} {\bibfnamefont {A.~I.}\ \bibnamefont {Kuznetsov}},
  \bibinfo {author} {\bibfnamefont {A.~E.}\ \bibnamefont {Miroshnichenko}},
  \bibinfo {author} {\bibfnamefont {Y.~F.}\ \bibnamefont {Yu}}, \ and\ \bibinfo
  {author} {\bibfnamefont {B.}~\bibnamefont {Luk’yanchuk}},\ }\href {\doibase
  10.1038/ncomms2538} {\bibfield  {journal} {\bibinfo  {journal} {Nat.
  Commun.}\ }\textbf {\bibinfo {volume} {4}},\ \bibinfo {pages} {1527}
  (\bibinfo {year} {2013})}\BibitemShut {NoStop}%
\bibitem [{\citenamefont {Vecchi}\ \emph {et~al.}(2009)\citenamefont {Vecchi},
  \citenamefont {Giannini},\ and\ \citenamefont
  {G\'omez~Rivas}}]{PhysRevLett.102.146807}%
  \BibitemOpen
  \bibfield  {author} {\bibinfo {author} {\bibfnamefont {G.}~\bibnamefont
  {Vecchi}}, \bibinfo {author} {\bibfnamefont {V.}~\bibnamefont {Giannini}}, \
  and\ \bibinfo {author} {\bibfnamefont {J.}~\bibnamefont {G\'omez~Rivas}},\
  }\href {\doibase 10.1103/PhysRevLett.102.146807} {\bibfield  {journal}
  {\bibinfo  {journal} {Phys. Rev. Lett.}\ }\textbf {\bibinfo {volume} {102}},\
  \bibinfo {pages} {146807} (\bibinfo {year} {2009})}\BibitemShut {NoStop}%
\bibitem [{\citenamefont {Esteban}\ \emph {et~al.}(2010)\citenamefont
  {Esteban}, \citenamefont {Teperik},\ and\ \citenamefont
  {Greffet}}]{PhysRevLett.104.026802}%
  \BibitemOpen
  \bibfield  {author} {\bibinfo {author} {\bibfnamefont {R.}~\bibnamefont
  {Esteban}}, \bibinfo {author} {\bibfnamefont {T.~V.}\ \bibnamefont
  {Teperik}}, \ and\ \bibinfo {author} {\bibfnamefont {J.~J.}\ \bibnamefont
  {Greffet}},\ }\href {\doibase 10.1103/PhysRevLett.104.026802} {\bibfield
  {journal} {\bibinfo  {journal} {Phys. Rev. Lett.}\ }\textbf {\bibinfo
  {volume} {104}},\ \bibinfo {pages} {026802} (\bibinfo {year}
  {2010})}\BibitemShut {NoStop}%
\bibitem [{\citenamefont {Caldarola}\ \emph {et~al.}(2015)\citenamefont
  {Caldarola}, \citenamefont {Albella}, \citenamefont {Cortés}, \citenamefont
  {Rahmani}, \citenamefont {Roschuk}, \citenamefont {Grinblat}, \citenamefont
  {Oulton}, \citenamefont {Bragas},\ and\ \citenamefont {Maier}}]{ncomms8915}%
  \BibitemOpen
  \bibfield  {author} {\bibinfo {author} {\bibfnamefont {M.}~\bibnamefont
  {Caldarola}}, \bibinfo {author} {\bibfnamefont {P.}~\bibnamefont {Albella}},
  \bibinfo {author} {\bibfnamefont {E.}~\bibnamefont {Cortés}}, \bibinfo
  {author} {\bibfnamefont {M.}~\bibnamefont {Rahmani}}, \bibinfo {author}
  {\bibfnamefont {T.}~\bibnamefont {Roschuk}}, \bibinfo {author} {\bibfnamefont
  {G.}~\bibnamefont {Grinblat}}, \bibinfo {author} {\bibfnamefont {R.~F.}\
  \bibnamefont {Oulton}}, \bibinfo {author} {\bibfnamefont {A.~V.}\
  \bibnamefont {Bragas}}, \ and\ \bibinfo {author} {\bibfnamefont {S.~A.}\
  \bibnamefont {Maier}},\ }\href {\doibase 10.1038/ncomms8915} {\bibfield
  {journal} {\bibinfo  {journal} {Nat. Commun.}\ }\textbf {\bibinfo {volume}
  {6}},\ \bibinfo {pages} {7915} (\bibinfo {year} {2015})}\BibitemShut
  {NoStop}%
\bibitem [{\citenamefont {Lin}\ and\ \citenamefont {Zheng}(2015)}]{srep14788}%
  \BibitemOpen
  \bibfield  {author} {\bibinfo {author} {\bibfnamefont {L.}~\bibnamefont
  {Lin}}\ and\ \bibinfo {author} {\bibfnamefont {Y.}~\bibnamefont {Zheng}},\
  }\href {\doibase 10.1038/srep14788} {\bibfield  {journal} {\bibinfo
  {journal} {Sci. Rep.}\ }\textbf {\bibinfo {volume} {5}},\ \bibinfo {pages}
  {14788} (\bibinfo {year} {2015})}\BibitemShut {NoStop}%
\bibitem [{\citenamefont {Kuhn}\ \emph {et~al.}(2017)\citenamefont {Kuhn},
  \citenamefont {Kosloff}, \citenamefont {Stickler}, \citenamefont {Patolsky},
  \citenamefont {Hornberger}, \citenamefont {Arndt},\ and\ \citenamefont
  {Millen}}]{Stefan17}%
  \BibitemOpen
  \bibfield  {author} {\bibinfo {author} {\bibfnamefont {S.}~\bibnamefont
  {Kuhn}}, \bibinfo {author} {\bibfnamefont {A.}~\bibnamefont {Kosloff}},
  \bibinfo {author} {\bibfnamefont {B.~A.}\ \bibnamefont {Stickler}}, \bibinfo
  {author} {\bibfnamefont {F.}~\bibnamefont {Patolsky}}, \bibinfo {author}
  {\bibfnamefont {K.}~\bibnamefont {Hornberger}}, \bibinfo {author}
  {\bibfnamefont {M.}~\bibnamefont {Arndt}}, \ and\ \bibinfo {author}
  {\bibfnamefont {J.}~\bibnamefont {Millen}},\ }\href
  {http://www.osapublishing.org/optica/abstract.cfm?URI=optica-4-3-356}
  {\bibfield  {journal} {\bibinfo  {journal} {Optica}\ }\textbf {\bibinfo
  {volume} {4}},\ \bibinfo {pages} {356} (\bibinfo {year} {2017})}\BibitemShut
  {NoStop}%
\bibitem [{\citenamefont {Fourkas}(2001)}]{Fourkas:01}%
  \BibitemOpen
  \bibfield  {author} {\bibinfo {author} {\bibfnamefont {J.~T.}\ \bibnamefont
  {Fourkas}},\ }\href {\doibase 10.1364/OL.26.000211} {\bibfield  {journal}
  {\bibinfo  {journal} {Opt. Lett.}\ }\textbf {\bibinfo {volume} {26}},\
  \bibinfo {pages} {211} (\bibinfo {year} {2001})}\BibitemShut {NoStop}%
\bibitem [{\citenamefont {Zhanghao}\ \emph {et~al.}(2019)\citenamefont
  {Zhanghao}, \citenamefont {Chen}, \citenamefont {Liu}, \citenamefont {Li},
  \citenamefont {Liu}, \citenamefont {Wang}, \citenamefont {Luo}, \citenamefont
  {Wang}, \citenamefont {Shan}, \citenamefont {Xie}, \citenamefont {Gao},
  \citenamefont {Chen}, \citenamefont {Jin}, \citenamefont {Li}, \citenamefont
  {Zhang}, \citenamefont {Dai},\ and\ \citenamefont {Xi}}]{nc104694}%
  \BibitemOpen
  \bibfield  {author} {\bibinfo {author} {\bibfnamefont {K.}~\bibnamefont
  {Zhanghao}}, \bibinfo {author} {\bibfnamefont {X.}~\bibnamefont {Chen}},
  \bibinfo {author} {\bibfnamefont {W.}~\bibnamefont {Liu}}, \bibinfo {author}
  {\bibfnamefont {M.}~\bibnamefont {Li}}, \bibinfo {author} {\bibfnamefont
  {Y.}~\bibnamefont {Liu}}, \bibinfo {author} {\bibfnamefont {Y.}~\bibnamefont
  {Wang}}, \bibinfo {author} {\bibfnamefont {S.}~\bibnamefont {Luo}}, \bibinfo
  {author} {\bibfnamefont {X.}~\bibnamefont {Wang}}, \bibinfo {author}
  {\bibfnamefont {C.}~\bibnamefont {Shan}}, \bibinfo {author} {\bibfnamefont
  {H.}~\bibnamefont {Xie}}, \bibinfo {author} {\bibfnamefont {J.}~\bibnamefont
  {Gao}}, \bibinfo {author} {\bibfnamefont {X.}~\bibnamefont {Chen}}, \bibinfo
  {author} {\bibfnamefont {D.}~\bibnamefont {Jin}}, \bibinfo {author}
  {\bibfnamefont {X.}~\bibnamefont {Li}}, \bibinfo {author} {\bibfnamefont
  {Y.}~\bibnamefont {Zhang}}, \bibinfo {author} {\bibfnamefont
  {Q.}~\bibnamefont {Dai}}, \ and\ \bibinfo {author} {\bibfnamefont
  {P.}~\bibnamefont {Xi}},\ }\href {\doibase 10.1038/s41467-019-12681-w}
  {\bibfield  {journal} {\bibinfo  {journal} {Nat. Commun.}\ }\textbf {\bibinfo
  {volume} {10}},\ \bibinfo {pages} {4694} (\bibinfo {year}
  {2019})}\BibitemShut {NoStop}%
\bibitem [{\citenamefont {Lethiec}\ \emph {et~al.}(2014)\citenamefont
  {Lethiec}, \citenamefont {Laverdant}, \citenamefont {Vallon}, \citenamefont
  {Javaux}, \citenamefont {Dubertret}, \citenamefont {Frigerio}, \citenamefont
  {Schwob}, \citenamefont {Coolen},\ and\ \citenamefont
  {Ma\^{\i}tre}}]{PhysRevX.4.021037}%
  \BibitemOpen
  \bibfield  {author} {\bibinfo {author} {\bibfnamefont {C.}~\bibnamefont
  {Lethiec}}, \bibinfo {author} {\bibfnamefont {J.}~\bibnamefont {Laverdant}},
  \bibinfo {author} {\bibfnamefont {H.}~\bibnamefont {Vallon}}, \bibinfo
  {author} {\bibfnamefont {C.}~\bibnamefont {Javaux}}, \bibinfo {author}
  {\bibfnamefont {B.}~\bibnamefont {Dubertret}}, \bibinfo {author}
  {\bibfnamefont {J.-M.}\ \bibnamefont {Frigerio}}, \bibinfo {author}
  {\bibfnamefont {C.}~\bibnamefont {Schwob}}, \bibinfo {author} {\bibfnamefont
  {L.}~\bibnamefont {Coolen}}, \ and\ \bibinfo {author} {\bibfnamefont
  {A.}~\bibnamefont {Ma\^{\i}tre}},\ }\href {\doibase
  10.1103/PhysRevX.4.021037} {\bibfield  {journal} {\bibinfo  {journal} {Phys.
  Rev. X}\ }\textbf {\bibinfo {volume} {4}},\ \bibinfo {pages} {021037}
  (\bibinfo {year} {2014})}\BibitemShut {NoStop}%
\bibitem [{\citenamefont {Sepioł}\ \emph {et~al.}(1997)\citenamefont
  {Sepioł}, \citenamefont {Jasny}, \citenamefont {Keller},\ and\ \citenamefont
  {Wild}}]{SEPIOL1997444}%
  \BibitemOpen
  \bibfield  {author} {\bibinfo {author} {\bibfnamefont {J.}~\bibnamefont
  {Sepioł}}, \bibinfo {author} {\bibfnamefont {J.}~\bibnamefont {Jasny}},
  \bibinfo {author} {\bibfnamefont {J.}~\bibnamefont {Keller}}, \ and\ \bibinfo
  {author} {\bibfnamefont {U.~P.}\ \bibnamefont {Wild}},\ }\href {\doibase
  https://doi.org/10.1016/S0009-2614(97)00622-2} {\bibfield  {journal}
  {\bibinfo  {journal} {Chem. Phys. Lett.}\ }\textbf {\bibinfo {volume}
  {273}},\ \bibinfo {pages} {444} (\bibinfo {year} {1997})}\BibitemShut
  {NoStop}%
\bibitem [{\citenamefont {Bartko}\ and\ \citenamefont
  {Dickson}(1999{\natexlab{a}})}]{doi:10.1021/jp993364q}%
  \BibitemOpen
  \bibfield  {author} {\bibinfo {author} {\bibfnamefont {A.~P.}\ \bibnamefont
  {Bartko}}\ and\ \bibinfo {author} {\bibfnamefont {R.~M.}\ \bibnamefont
  {Dickson}},\ }\href {\doibase 10.1021/jp993364q} {\bibfield  {journal}
  {\bibinfo  {journal} {J. Phys. Chem. B}\ }\textbf {\bibinfo {volume} {103}},\
  \bibinfo {pages} {11237} (\bibinfo {year} {1999}{\natexlab{a}})}\BibitemShut
  {NoStop}%
\bibitem [{\citenamefont {Karedla}\ \emph {et~al.}(2015)\citenamefont
  {Karedla}, \citenamefont {Stein}, \citenamefont {H\"ahnel}, \citenamefont
  {Gregor}, \citenamefont {Chizhik},\ and\ \citenamefont
  {Enderlein}}]{PhysRevLett.115.173002}%
  \BibitemOpen
  \bibfield  {author} {\bibinfo {author} {\bibfnamefont {N.}~\bibnamefont
  {Karedla}}, \bibinfo {author} {\bibfnamefont {S.~C.}\ \bibnamefont {Stein}},
  \bibinfo {author} {\bibfnamefont {D.}~\bibnamefont {H\"ahnel}}, \bibinfo
  {author} {\bibfnamefont {I.}~\bibnamefont {Gregor}}, \bibinfo {author}
  {\bibfnamefont {A.}~\bibnamefont {Chizhik}}, \ and\ \bibinfo {author}
  {\bibfnamefont {J.}~\bibnamefont {Enderlein}},\ }\href {\doibase
  10.1103/PhysRevLett.115.173002} {\bibfield  {journal} {\bibinfo  {journal}
  {Phys. Rev. Lett.}\ }\textbf {\bibinfo {volume} {115}},\ \bibinfo {pages}
  {173002} (\bibinfo {year} {2015})}\BibitemShut {NoStop}%
\bibitem [{\citenamefont {Bartko}\ and\ \citenamefont
  {Dickson}(1999{\natexlab{b}})}]{doi:10.1021/jp9846330}%
  \BibitemOpen
  \bibfield  {author} {\bibinfo {author} {\bibfnamefont {A.~P.}\ \bibnamefont
  {Bartko}}\ and\ \bibinfo {author} {\bibfnamefont {R.~M.}\ \bibnamefont
  {Dickson}},\ }\href {\doibase 10.1021/jp9846330} {\bibfield  {journal}
  {\bibinfo  {journal} {J. Phys. Chem. B}\ }\textbf {\bibinfo {volume} {103}},\
  \bibinfo {pages} {3053} (\bibinfo {year} {1999}{\natexlab{b}})}\BibitemShut
  {NoStop}%
\bibitem [{\citenamefont {Lieb}\ \emph {et~al.}(2004)\citenamefont {Lieb},
  \citenamefont {Zavislan},\ and\ \citenamefont {Novotny}}]{Lieb:04}%
  \BibitemOpen
  \bibfield  {author} {\bibinfo {author} {\bibfnamefont {M.~A.}\ \bibnamefont
  {Lieb}}, \bibinfo {author} {\bibfnamefont {J.~M.}\ \bibnamefont {Zavislan}},
  \ and\ \bibinfo {author} {\bibfnamefont {L.}~\bibnamefont {Novotny}},\ }\href
  {\doibase 10.1364/JOSAB.21.001210} {\bibfield  {journal} {\bibinfo  {journal}
  {J. Opt. Soc. Am. B}\ }\textbf {\bibinfo {volume} {21}},\ \bibinfo {pages}
  {1210} (\bibinfo {year} {2004})}\BibitemShut {NoStop}%
\bibitem [{\citenamefont {Dodson}\ \emph {et~al.}(2014)\citenamefont {Dodson},
  \citenamefont {Kurvits}, \citenamefont {Li},\ and\ \citenamefont
  {Zia}}]{Dodson:14}%
  \BibitemOpen
  \bibfield  {author} {\bibinfo {author} {\bibfnamefont {C.~M.}\ \bibnamefont
  {Dodson}}, \bibinfo {author} {\bibfnamefont {J.~A.}\ \bibnamefont {Kurvits}},
  \bibinfo {author} {\bibfnamefont {D.}~\bibnamefont {Li}}, \ and\ \bibinfo
  {author} {\bibfnamefont {R.}~\bibnamefont {Zia}},\ }\href {\doibase
  10.1364/OL.39.003927} {\bibfield  {journal} {\bibinfo  {journal} {Opt.
  Lett.}\ }\textbf {\bibinfo {volume} {39}},\ \bibinfo {pages} {3927} (\bibinfo
  {year} {2014})}\BibitemShut {NoStop}%
\bibitem [{\citenamefont {Osorio}\ \emph {et~al.}(2015)\citenamefont {Osorio},
  \citenamefont {Mohtashami},\ and\ \citenamefont {Koenderink}}]{srep59966}%
  \BibitemOpen
  \bibfield  {author} {\bibinfo {author} {\bibfnamefont {C.~I.}\ \bibnamefont
  {Osorio}}, \bibinfo {author} {\bibfnamefont {A.}~\bibnamefont {Mohtashami}},
  \ and\ \bibinfo {author} {\bibfnamefont {A.~F.}\ \bibnamefont {Koenderink}},\
  }\href {\doibase 10.1038/srep09966} {\bibfield  {journal} {\bibinfo
  {journal} {Sci. Rep.}\ }\textbf {\bibinfo {volume} {5}},\ \bibinfo {pages}
  {9966} (\bibinfo {year} {2015})}\BibitemShut {NoStop}%
\bibitem [{\citenamefont {Curcio}\ \emph {et~al.}(2020)\citenamefont {Curcio},
  \citenamefont {Alemán-Castañeda}, \citenamefont {Brown}, \citenamefont
  {Brasselet},\ and\ \citenamefont {Alonso}}]{nc115307}%
  \BibitemOpen
  \bibfield  {author} {\bibinfo {author} {\bibfnamefont {V.}~\bibnamefont
  {Curcio}}, \bibinfo {author} {\bibfnamefont {L.~A.}\ \bibnamefont
  {Alemán-Castañeda}}, \bibinfo {author} {\bibfnamefont {T.~G.}\ \bibnamefont
  {Brown}}, \bibinfo {author} {\bibfnamefont {S.}~\bibnamefont {Brasselet}}, \
  and\ \bibinfo {author} {\bibfnamefont {M.~A.}\ \bibnamefont {Alonso}},\
  }\href {\doibase 10.1038/s41467-020-19064-6} {\bibfield  {journal} {\bibinfo
  {journal} {Nat. Commun.}\ }\textbf {\bibinfo {volume} {11}},\ \bibinfo
  {pages} {5307} (\bibinfo {year} {2020})}\BibitemShut {NoStop}%
\bibitem [{\citenamefont {Jin}\ \emph {et~al.}(2019)\citenamefont {Jin},
  \citenamefont {Yu},\ and\ \citenamefont {Zhang}}]{Jin:19}%
  \BibitemOpen
  \bibfield  {author} {\bibinfo {author} {\bibfnamefont {Y.}~\bibnamefont
  {Jin}}, \bibinfo {author} {\bibfnamefont {X.}~\bibnamefont {Yu}}, \ and\
  \bibinfo {author} {\bibfnamefont {J.}~\bibnamefont {Zhang}},\ }\href
  {\doibase 10.1364/JOSAB.36.002369} {\bibfield  {journal} {\bibinfo  {journal}
  {J. Opt. Soc. Am. B}\ }\textbf {\bibinfo {volume} {36}},\ \bibinfo {pages}
  {2369} (\bibinfo {year} {2019})}\BibitemShut {NoStop}%
\bibitem [{\citenamefont {Jin}\ \emph {et~al.}(2018)\citenamefont {Jin},
  \citenamefont {Yu},\ and\ \citenamefont {Zhang}}]{Jin:18}%
  \BibitemOpen
  \bibfield  {author} {\bibinfo {author} {\bibfnamefont {Y.}~\bibnamefont
  {Jin}}, \bibinfo {author} {\bibfnamefont {X.}~\bibnamefont {Yu}}, \ and\
  \bibinfo {author} {\bibfnamefont {J.}~\bibnamefont {Zhang}},\ }\href
  {\doibase 10.1007/s11433-018-9230-6} {\bibfield  {journal} {\bibinfo
  {journal} {Sci. China-Phys., Mech. Astron.}\ }\textbf {\bibinfo {volume}
  {61}},\ \bibinfo {pages} {114221} (\bibinfo {year} {2018})}\BibitemShut
  {NoStop}%
\end{thebibliography}%
\bibliographystyle{apsrev4-1}
\end{document}